\documentclass[aps,preprint,groupedaddress]{revtex4}
\usepackage{natbib,graphics,graphicx}

\begin{document}
\bibliographystyle{h-physrev}

\title{A Rejuvenated Universe Without Initial Singularity}

\author{Eric Gawiser}
\affiliation{Yale Astronomy Department, New Haven, CT  06520-8101}
\altaffiliation{NSF Astronomy \& Astrophysics Postdoctoral Fellow}
\affiliation{Departamento de Astronom\'{i}a, Universidad de Chile, 
Casilla 36-D, Las Condes, Santiago, Chile}
\altaffiliation{Andes Prize Fellow}

\date{\today}
\begin{abstract}
Cosmological observations strongly suggest the presence of dark
energy which comprises the majority of the current energy density 
of the universe.  The equation of state relating the pressure 
and energy density of this dark energy, $p = w \rho$, appears to have   
$w \simeq -1$, 
with most analyses preferring 
$w < -1$ when these values are considered as part of the parameter space.
If $w<-1$ is the future behavior of the dark energy, the scale factor, 
expansion rate, and energy density
 of the universe will diverge in a finite time as its 
apparent expansion age tends to zero.  
We hypothesize that $w>-1$ is restored at 
a late enough time, perhaps due to a phase transition of the dark energy, 
and show that this produces 
 conditions observationally indistinguishable from a Hot Big Bang.  
This process of rejuvenation may have occurred in the past, 
making our universe much older than it appears and eliminating the Big 
Bang singularity.  
\end{abstract}

\pacs{98.80.Jk,98.80.Bp}

\maketitle

The expansion of the universe follows the Friedmann equations.
The  
rate of change of the scale factor $a$ is given by 
\begin{equation}
H^2=\left(\frac{\dot{a}}{a}\right)^2=\frac{8 \pi G}{3} \rho - \frac{k}{a^2}
\;,\;\;\; 
\label{eq:friedmann1}
\end{equation}
where $H$ is the Hubble ``constant'', G is Newton's constant, $\rho$ 
is the total energy density, and $k$ is the spatial curvature.  
The scale factor is arbitrary up to a multiplicative constant 
but the fractional rate of expansion H 
is observable, leading to Hubble's law of recession $v=Hd$ 
 and an apparent expansion age of the 
universe given by $t_{expansion} = H^{-1}$.  
The second Friedmann equation illustrates the impact of pressure 
upon the acceleration or deceleration of the expansion, 
\begin{equation}
\frac{\ddot{a}}{a}=-\frac{4 \pi G}{3} \left( \rho + 3 p \right)
=-\frac{4 \pi G}{3} \rho \left( 1 + 3 w_{eff} \right)
\;.\;\;\; 
\label{eq:friedmann2}
\end{equation}
Hence the strong observational evidence for accelerating 
expansion \cite{perlmutteretal99,riessetal98} 
implies that the average equation of state of the radiation, matter, 
and dark energy in the universe has $w_{eff} < -1/3$.  
If no interchange is occurring between the matter and dark energy, 
they are separately conserved, and their energy densities are 
related to the scale factor by $\rho_m=\rho_{m,0}(a/a_0)^{-3}$ 
and $\rho_\phi=\rho_{\phi,0}(a/a_0)^{-3(1+w)}$.  

\begin{figure}
\scalebox{0.7}[0.7]{\rotatebox{0}{\includegraphics{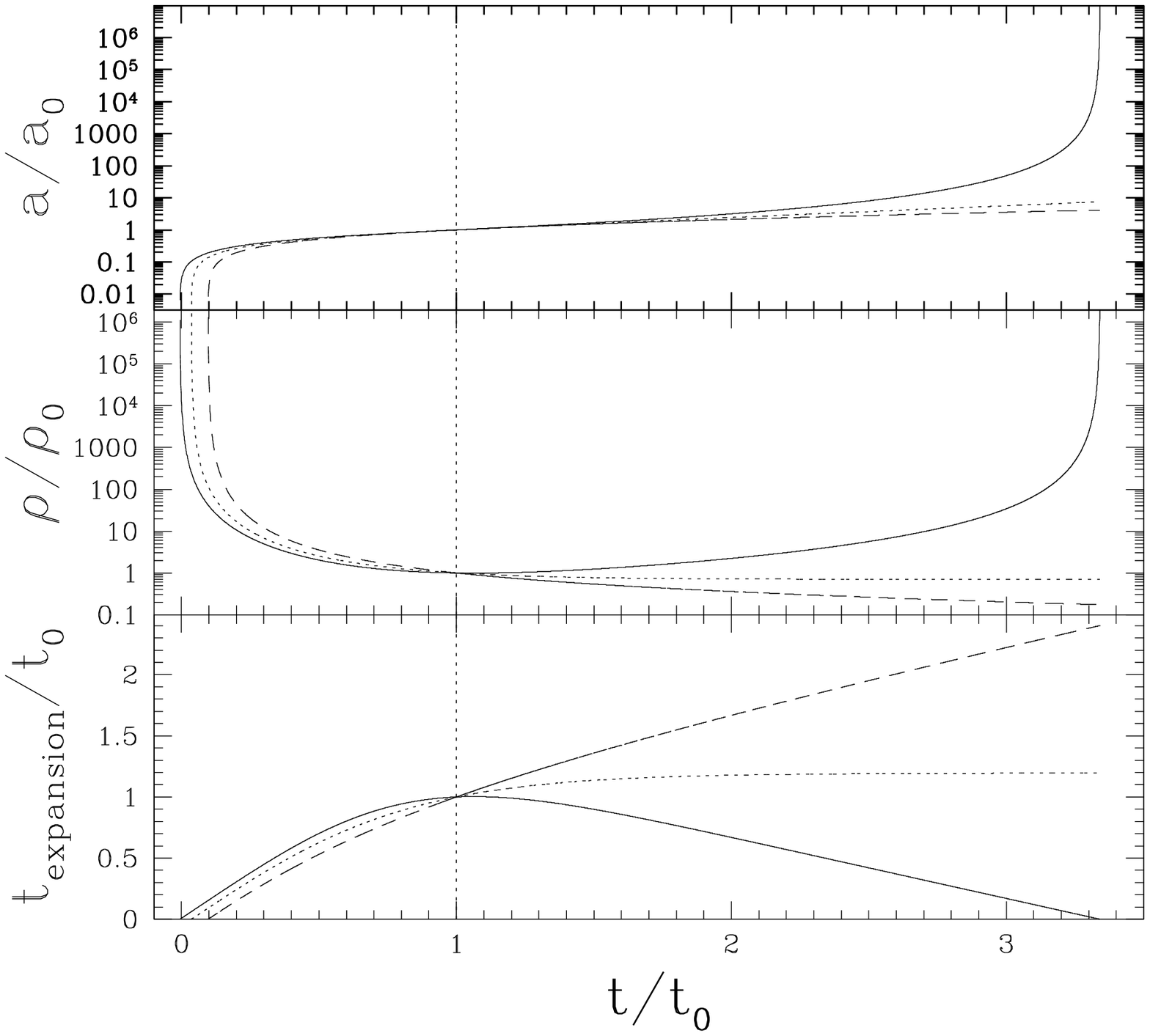}}}
\caption{\small 
The effects of dark energy with constant 
$w=-2/3$ (dashed), $w=-1$ (dotted), and $w=-4/3$ (solid).  Vertical dotted 
line indicates the present era.  Top panel shows 
the scale factor $a$ in units of the scale factor today versus
time in units of the current expansion age $t_0=H_0^{-1}$.  
The scale factor is growing exponentially for $w=-1$ but diverges 
when the universe reaches about 3.3 times its current expansion age 
for $w=-4/3$.  The middle panel shows that energy density in units of the 
current energy density becomes constant for $w=-1$ but diverges if $w=-4/3$.  
The bottom panel shows that the expansion age $H^{-1}$ 
becomes 
constant for $w=-1$ but decreases to zero for $w=-4/3$. 
If $w=-4/3$, the observable quantities of 
energy density and expansion age return to their Big Bang values as the 
scale factor diverges. 
}
\label{fig:3model}
\end{figure}

Observations show that our universe is nearly 
flat ($k=0$) and has dark energy with $w \simeq -1$ which 
dominates the current energy density \cite{bahcalletal99,wangetal00,
garnavichetal98b,perlmuttertw99,
wagam99,sainietal00,knopetal03,spergeletal03,melchiorrietal03}.  
Most analyses prefer 
$w < -1$, dubbed 
 phantom energy \cite{caldwell02}, 
when these values are considered as part of the parameter space, 
although $w \geq -1$ is often assumed as a theoretical prior.
We can set the scale factor today $a_0=1$ and 
specialize (\ref{eq:friedmann1}) to analyze the past 
and future expansion rate of the universe in units of the current 
Hubble constant $H_0\approx70$km/s/Mpc \cite{freedman00}, resulting in 
a simple differential equation,  
\begin{equation}
\dot{a}=( \Omega_{m,0}a^{-1} + \Omega_{\phi ,0} a^{-1-3w})^{1/2}
\;,\;\;\; 
\label{eq:friedmann3}
\end{equation}
where WMAP determined $\Omega_{m,0}=8 \pi G \rho_{m,0} / (3 H_0^2) = 0.26$ 
and $\Omega_{\phi,0} = 0.74$.   
Figure \ref{fig:3model} illustrates the behavior of the scale factor, 
energy density, and expansion age of the universe for the cases 
$w=-2/3, -1, -4/3$.  
 In each of the 
models, our Universe appears to have begun in a Big Bang when the scale 
factor and expansion age were equal to zero and 
the energy density was infinite.  This initial singularity occurs at a 
slightly different time in the three models, which reflects the amount by 
which 
the true age of the universe today in each model differs from the 
current expansion age of 14 billion years.  
Allowing $w<-1$ violates the dominant energy condition of general 
relativity and allows the energy content of the dark energy 
to flow faster than the speed of light \cite{wald84}.  This dominant 
energy condition is indeed violated by some proposed models, although 
care must be taken to avoid ``tachyonic'' modes with negative 
squared effective mass that are unstable to 
growth \cite{parkerr01,barrow88,pollock88,caldwell02,schulzw01}.  
A thorough investigation by \cite{carrollht03} indicates that 
viable models with $w<-1$ can be constructed 
but require significant care.  
$w<-1$ is equivalent to 
an effective cosmological constant that increases with time, 
and it is possible that string physics \cite{frampton03} or 
other extra-dimensional effects could give the dark energy an effective 
$w<-1$ without creating instabilities.  
Moreover, the phenomenology described by $w<-1$ could instead be achieved 
with a pure cosmological constant and an increasing value of Newton's 
gravitational constant $G$ (see \cite{barrow99}).  
With the running of coupling constants intrinsic to M-theory and 
the recent barrage of models with large extra dimensions, it 
therefore seems 
premature to dismiss the idea that the dark energy has $w_{eff}<-1$ 
as unphysical.  

The behavior of the $w=-4/3$ model is generic 
for all models with $w<-1$ even if $w$ is not constant; the universe 
undergoes superinflation 
where the scale factor diverges within finite 
time as the energy density becomes infinite and the expansion age 
reaches zero.  Let $H_d$ be the expansion rate at a time $t_d$ where 
matter has become irrelevant and dark energy dominates the energy density.
Then if the equation of state is constant with $w<-1$, 
for $t > t_d$ (\ref{eq:friedmann1}) 
has the solution \cite{lucchinm85} 
\begin{equation}
a=a_d \left( 1 - \frac{3}{2} (-1-w)H_d(t-t_d)\right)^{-\frac{2}{3(-1-w)}}
=a_d \left( \frac{3}{2} (-1-w)H_d(t_\infty-t)\right)^{-\frac{2}{3(-1-w)}}
\;,\;\;\; 
\label{eq:friedmann4}
\end{equation}
where the time of scale factor divergence $t_\infty$ is given by 
\cite{colesl95} 
\begin{equation}
t_\infty=t_d + \frac{2}{3(-1-w)H_d}
\;.\;\;\; 
\label{eq:divergence}
\end{equation}
The conformal time is well behaved, approaching a finite positive 
value as $t \rightarrow t_\infty$.  
For $w<-1$, the 
physical observables of energy density and expansion age at divergence 
are identical 
to those that exist at the initial moment of the Big Bang cosmology, with 
the notable exception that the energy density is in the form of dark 
energy whose behavior is closer to a cosmological constant than to 
matter or radiation.  The moment of divergent scale factor represents 
a singularity, and observers in this superinflationary universe 
have event horizons that shrink to zero size at the moment the scale factor 
diverges because even nearby objects are receding faster than the 
speed of light.  While $H(t)$ is observable, 
determining the true age of the universe requires 
a full knowledge of the historical equation of state of each component 
of matter, radiation, and dark energy.  Age determinations from 
galaxy recession velocities (the expansion age), stellar evolution,  
 and nuclear isotope abundances will differ 
from each other in a stage of dark energy domination as the 
expansion age differs greatly from the true age of the universe.  

\begin{figure}
\scalebox{0.7}[0.7]{\rotatebox{0}{\includegraphics{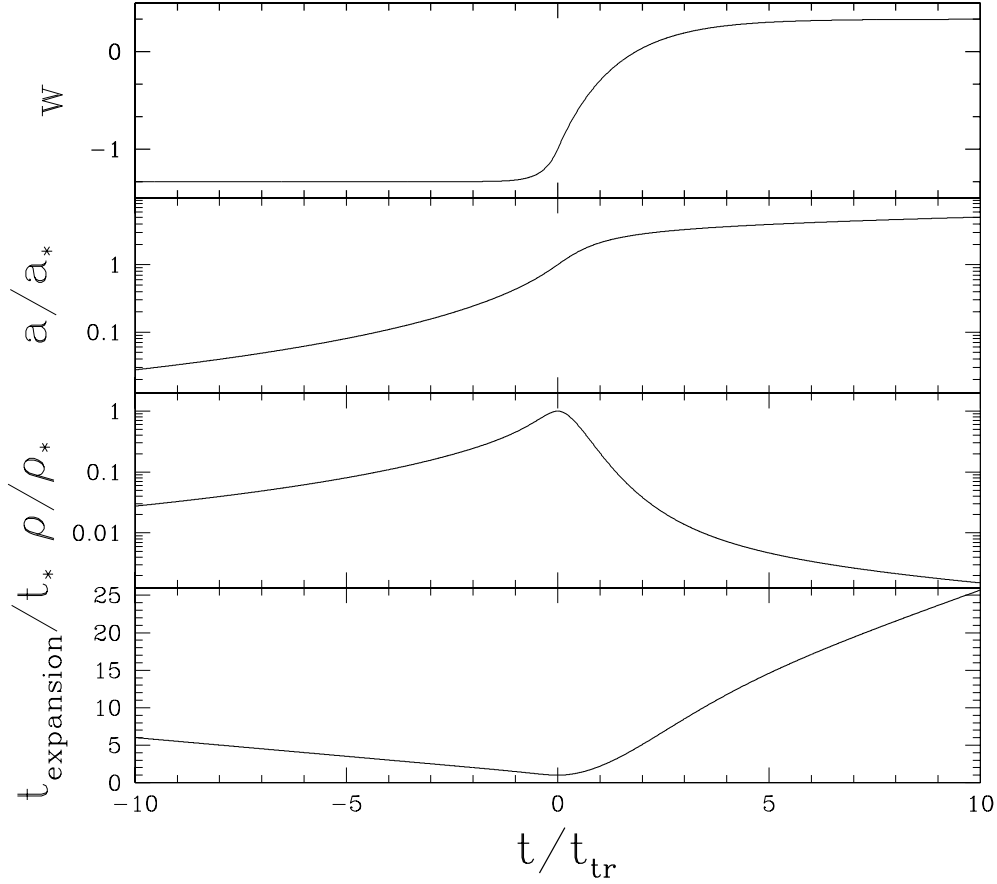}}}
\caption{\small 
A toy model illustrating how a final singularity can be avoided by  
changing the equation of state of the dark energy
before the scale factor diverges.  
The characteristic timescale for this transition is $t_{tr}$ 
which might correspond to a 
Planck time of $10^{-43}$ seconds.  The top panel shows $w$ undergoing 
a transition from $-4/3$ (phantom 
energy) to $1/3$ (radiation) with the closest approach to a singularity 
at the moment when $w=-1$ marked as $t=0$.  
The scale factor (second panel) 
expands throughout the entire transition but the 
acceleration of the expansion ceases.  The energy density (third panel)
reaches its maximum value $\rho_*$ at $t=0$ and then begins to decrease.  
The apparent age of the universe, its expansion age (bottom panel), 
reaches a minimum of $t_*$ at $t=0$.  }
\label{fig:bigbang}
 \end{figure}

As the moment of divergence approaches, the energy density of the universe 
reaches scales at which physics beyond the 
standard model may alter the behavior.  Figure \ref{fig:bigbang} shows a 
toy model in which the dark energy undergoes a phase transition from 
$w=-4/3$ to $w=1/3$.  A final singularity is avoided, and a phase of 
decelerating expansion begins which looks just like the Big Bang but 
without an initial singularity.  The singularity theorems of general 
relativity do not apply because they presume that the strong energy 
condition 
($\rho + p > 0$) holds \cite {wald84} but this is not applicable for 
$\rho>0, w<-1$. 
Hence a runaway universe that undergoes this 
sort of phase transition ends 
up rejuvenated with a small expansion age and re-energized with a 
correspondingly large energy density.  

To create a Hot Big Bang in which 
nucleosynthesis and the cosmic background radiation are consistent with 
observations, 
the phase transition must produce $w=1/3$ 
at the end and then generate matter through thermal production.  
An intermediate step could be to produce a 
$w=0$ (matter) component that decays into 
radiation ($w=1/3$) as at the end of the standard inflationary scenario 
\cite{guth81,linde82,albrechts82,guth00,linde00}.
One possible mechanism to generate matter from the dark energy 
would be 
gravitational particle production into 
nonrelativistic particles of mass $M_X$, 
which occurs once $\dot{H}/H > M_X$ \cite{chungetal01}
and is therefore guaranteed 
to happen at some point in the approach to scale factor divergence 
regardless of the $M_X$ available.  A radiation-dominated Hot Big Bang
would then be produced by particle decay.  However, it seems impossible to get 
enough of the energy density into matter by this mechanism 
to make  
$w_{eff} >-1$  
because the gravitational particle production ends 
as $w_{eff} \rightarrow -1$ and 
$\dot{H}/H \rightarrow < M_X$.  Decay of the matter ($w=0$) into radiation 
($w=1/3$) 
could then make $w_{eff}>-1$ but could not achieve $w_{eff}>-1/3$ which 
is required to end the accelerating expansion and prevent a return to an 
epoch of dark energy domination.  
Another possible mechanism would be for the kinetic energy 
of a scalar phantom energy field to come to dominate its potential; 
since $w = (\frac{1}{2} \dot{\phi}^2 - V)/(\frac{1}{2} \dot{\phi}^2 + V)$, 
kinetic energy domination produces $w=1$.  

For a flat universe, the relationship between $H$ and $\rho$ is determined 
entirely by (\ref{eq:friedmann1}) 
so rejuvenation leads to the precise physical observables 
which characterize the Big Bang.  The runaway expansion of a 
$w<-1$ universe is guaranteed to reduce any curvature tremendously because 
of its $a^{-2}$ dependence, solving the flatness problem.  
The horizon problem of the Big 
Bang model is also resolved as regions initially under causal contact 
are expanded to very large, apparently superhorizon, size before $t=0$.  
The presence of event horizons in the rejuvenated universe depends upon 
the relative length of decelerating and accelerating expansion epochs.  
The monopole 
problem is of interest as any massive remnants produced at the maximum 
energy density will remain today unless the equation of state of the 
universe remains near $w=-1$ 
for long enough to reduce their abundance significantly.  The precise 
spectrum of density perturbations generated in a rejuvenated universe 
will depend upon the details of the dark energy and its phase transition, 
but a stage resembling the standard 
inflationary scenario 
is necessary to 
produce a nearly scale-invariant spectrum of density fluctuations consistent 
with observations of cosmological structure.  
There are interesting similarities between the rejuvenation scenario and 
the model of Pre-Big-Bang inflation based upon the scale factor 
duality of string theory \cite{veneziano97,gasperiniv93}.  
However, the PBB model has growing curvature and, 
for three spatial dimensions, has $w=-1/3$ at $t<0$.  Recent attempts to 
avoid a Big Bang singularity such as PBB or the Ekpyrotic Universe 
\cite{khouryetal01,khouryetal02a} 
appeal to flat, static spacetime as an ideal set of initial conditions, 
although fine-tuning of initial curvature appears necessary.    
Rejuvenation instead uses the current observed 
state of our universe as initial conditions, 
with added assumptions about the equation of state of the dark energy 
and its future behavior.  

This discussion of the rejuvenation scenario leads to three separate 
hypotheses, each of which makes predictions that can be tested against 
cosmological observations.  The first is that rejuvenation will occur 
in the future of our universe; this requires $w<-1$ for a long enough 
period to increase the energy density to levels expected in the 
first second of the Hot Big Bang model.  While this does not technically 
require $w<-1$ or $dw/dt < 0$ today, future rejuvenation will seem 
extremely improbable if neither of those conditions is met by the current 
dark energy.  There are observational approaches using Type Ia supernovae, 
gravitational lensing, and  
the evolution of the number abundance of dark matter halos 
that should reveal if the present dark energy has $w<-1$ or
 $dw/dt<0$.\cite{efstathiou99,cappi01,newmand00,huterert01}  
Unfortunately, neither of those conditions would guarantee a sufficiently long 
period of $w<-1$ to allow rejuvenation; proof of that would
require a fundamental understanding of the nature of the dark 
energy. 

If the dark energy has a low sound speed, 
existing density inhomogeneities allow the runaway universe to 
provide a foundation for eternal (chaotic) inflation \cite{linde00,guth00} 
as regions with lower than average mass 
density today already have higher than average expansion rates and 
will undergo rejuvenation sooner than average.  If the dark energy 
equation-of-state is constant throughout space, 
regions that are 
gravitationally bound in the current universe will never achieve dark 
energy domination as they have already decoupled from the universal 
expansion.    
Thus density fluctuations become destiny fluctuations, and the edge of 
bound regions becomes an event horizon at the moment of scale factor 
divergence if rejuvenation does not occur.   
If, on the other hand, the sound speed of the dark energy 
is the speed of light as is typical 
for scalar-field ``quintessence'' models, the dark energy density will 
roughly equilibrate (this precludes a universally constant value of 
$w$) and current gravitationally bound objects will be ripped apart - this 
has been dubbed ``Cosmic Doomsday'' by \cite{caldwellkw03}.  
Note that current density fluctuations will prevent even this fate from 
being synchronized on scales of order the horizon size.  
Would black holes from our current universe survive rejuvenation?  It appears 
so, but like any pre-inflationary relic their density will be diluted 
by the expansion to the point where observers in the post-rejuvenation 
phase would be quite unlikely to find them.   

The second hypothesis is that rejuvenation generated the hot early 
conditions of our universe, avoiding a Big Bang singularity and 
making our universe much older than it appears.  
The apparent age of the universe determined from its expansion 
dynamics, content of matter and dark energy, and from nucleosynthetic 
ages will all be misleading in this case.  
Evidence that the fundamental physics of the 
dark energy allows a direct connection between the dark 
energy of a previous super-inflationary phase 
and the dark energy today would make our second hypothesis  
easier to believe, but this connection is not necessary for the hypothesis 
to be correct.
Rejuvenation predicts that the amplitude of primordial density fluctuations 
should be increasing with decreasing physical scale i.e. 
the scalar spectral index $n=d \log P(k) / d \log k >1$ over 
a range of physical scales that exited the horizon while the expansion 
was still accelerating.  
This provides an intriguing explanation for the anomalously 
low values of the quadrupole and octopole in the WMAP skymaps
\cite{bennettetal03}.  
It appears possible for rejuvenation models to generate primordial 
power spectra that run from blue to red i.e. $n>1$ to $n<1$ over the range 
of scales probed by current observations of cosmological structure, as 
preferred by 
recent WMAP results \cite{peirisetal03}.  These observable perturbations 
were generated about 50 e-foldings before the end of inflation or 
rejuvenation \cite{liddlel03}, although the number of e-folds 
can be smaller in the latter case.  
The standard slow-roll inflationary perturbation solution 
yields a formally infinite contribution to the primordial power spectrum
at the scale corresponding to the horizon size 
when $w$ crosses through the value $-1$.
Forming a viable cosmological model of rejuvenation will require careful 
attention to this problem, which could be resolved by ending slow-roll early 
in the phase transition.
 The detailed physics of cosmological perturbations generated 
during rejuvenation will be addressed in future work.   

The third hypothesis is that we live in 
``Phoenix'' 
universe which undergoes successive 
stages of rejuvenation, becoming larger and even closer to flat each time.  
This contrasts with all
 previous models of this type which have oscillated between Big Bang and 
Big Crunch, including the recently proposed Cyclic Universe  
\cite{steinhardtt02a}.    
Beyond requiring the first and second hypotheses to be true and 
thereby echoing their predictions, this third hypothesis might best 
be relegated to the realm of metaphysics, as the current universe likely 
offers few clues to the possible existence of multiple rejuvenation 
phases in the past or future.  Nonetheless, a full physical understanding 
of the dark energy could illuminate an oscillatory nature, 
perhaps due to a scalar field endlessly rolling 
up and then falling back down its potential -- a ``Sisyphus'' universe.    
Rejuvenation does not, however, 
allow for a periodic universe where the future actually produces the past.  
The monotonically increasing scale factor makes the 
universe much larger at a given approach to singularity than at 
the previous approach, with increased entropy and density inhomogeneities.  
The presence or absence of event horizons, which are anathema to M-theory, 
would depend on the relative rate of expansion during accelerating 
and decelerating periods back to the infinite past and out to the infinite 
future; this is an issue worth exploring in more detail for both rejuvenation 
and the Cyclic Universe models.  

It should be noted that there is still 
no satisfactory explanation for the presence of dark energy today.
One can postulate that the decay 
of the dark energy of the previous inflationary phase 
into matter and radiation was incomplete, but details 
await a physical theory that accounts for the dark energy and any 
phase transitions it may undergo.  A particular challenge is 
to explain the time-asymmetry of the rejuvenation phase transition i.e. 
why $w$ remains greater than $-1$ as the universe cools to the temperature 
where $w<-1$ occurred previously.    
Pending a well-defined physical theory for the dark energy, 
the rejuvenated universe scenario 
provides a plausible connection between 
the mystery of the present dark energy and 
 the initial conditions for the inflationary phase of the 
``early'' universe.   
It remains a significant observational challenge
 to determine whether our 
universe originated in a Big Bang singularity or whether phantom 
energy provided it with a fountain of youth. 

\begin{acknowledgments}
I acknowledge valuable conversations with Paolo Gondolo, 
Kim Griest, Andrew Liddle, Sevil Salur, David Spergel, and Martin White 
and financial support from Fundaci{\'o}n Andes and an NSF 
Astronomy \& Astrophysics Postdoctoral Fellowship (AAPF) through 
grant AST-0201667.  
\end{acknowledgments}

%\bibliography{refs.bib}

%\newpage

%{\bf ~}
%Random text here was needed or it ignored figures.

\end{document}